\documentclass{aa}
\usepackage{natbib}
\usepackage{epsfig}
\def\ut#1{\mathop{\vtop{\ialign{##\crcr
     $\hfil\displaystyle{#1}\hfil$\crcr\noalign
     {\kern1pt\nointerlineskip}\hbox{$\hfil\sim\hfil$}\crcr
     \noalign{\kern1pt}}}}}

\def\undersymbol#1#2{\mathop{\vtop{\ialign{##\crcr
     $\hfil\displaystyle{#2}\hfil$\crcr\noalign
     {\kern1pt\nointerlineskip}\hbox{$\hfil#1\hfil$}\crcr
     \noalign{\kern1pt}}}}}
\def\arcsec{^{\prime\prime}}
\def\arcmin{^{\prime}}
\def\degr{^0}

\usepackage{graphicx}

\begin{document}

\title{Messier 81's Planck view vs  its halo mapping}
       \author{V.G. Gurzadyan\inst{1,2},
       F. De Paolis\inst{3,4}, A.A. Nucita\inst{3,4},   A.L. Kashin\inst{1}, A. Amekhyan\inst{1}, S. Sargsyan\inst{1}, G. Yegorian\inst{1}, A. Qadir\inst{5}, G. Ingrosso\inst{3,4}, Ph. Jetzer\inst{6},   \and D. Vetrugno\inst{7}}
       
              \institute{Center for Cosmology and Astrophysics, Alikhanian National Laboratory and Yerevan State University, Yerevan, Armenia \and
                               SIA, Sapienza University of Rome, Rome, Italy               \and
              Dipartimento di Matematica e  Fisica ``E. De Giorgi'', Universit\`a del Salento, Via per Arnesano, I-73100, Lecce, Italy  
               \and INFN, Sezione  di Lecce, Via per Arnesano, I-73100, Lecce, Italy  
             \and
Department of Physics, School of Natural Sciences,
National University of Sciences and Technology, Islamabad, Pakistan
              \and
Physik-Institut, Universit\"at
Z\"urich, Winterthurerstrasse 190, 8057 Z\"urich, Switzerland
\and 
Department of Physics, University of Trento, I-38123 Povo, Trento, Italy and 
TIFPA/INFN, I-38123 Povo,  Italy
}

   \offprints{F. De Paolis, \email{depaolis@le.infn.it}}
   \date{Submitted: XXX; Accepted: XXX}

 \abstract{
 This paper is a follow-up of a previous paper about the M82 galaxy and its halo based on {\it Planck} observations. As in the case of M82, so also for the M81 galaxy a substantial North-South and East-West temperature asymmetry is found, extending up to galactocentric distances of about $1.5\degr$. The temperature asymmetry is almost frequency independent and can be interpreted as a Doppler-induced effect related to the M81 halo rotation and/or  triggered by the gravitational interaction of the galaxies within the M81  Group. Along with the analogous study of several nearby edge-on spiral galaxies, the CMB temperature asymmetry method thus is shown to act as a direct  tool to map the galactic haloes and/or the intergalactic bridges, invisible in other bands or by other methods. 
 }


   \authorrunning{Gurzadyan et al.}
   \titlerunning{Messier 81's {\it Planck} view vs its halo mapping}
   \maketitle
%

\section{Introduction}
Cosmic Microwave Background (CMB) data
are mainly used with the primary aim to infer the
values of the parameters of the cosmological standard model. In addition, CMB data also offer  a
unique opportunity to study the large-scale temperature asymmetries far beyond the size typically accessible with other tools toward nearby
astronomical systems (see, e.g., \citealt{rg,depaolis2011}). Here we continue the use of CMB data to map the dark haloes of nearby galaxies, the latter often studied in most details on other bands or via other methods.
Indeed, in the recent past we have analyzed {\it Planck} data toward four nearby galaxies with the main aim of testing if microwave data show a substantial temperature asymmetry of one side with respect to the other about  the rotation axis of the galactic disks. We have considered, in particular: M31 galaxy and its halo (\citealt{depaolis2014}) the active radio galaxy Centaurus A (Cen A)
that is considered the closest AGN
(\citealt{depaolis2015}),  M82, the largest galaxy in
the M81 Group in the Ursa Major constellation (\citealt{gurzadyan2015}), the M 33 galaxy where we found a substantial temperature asymmetry with respect to its minor axis projected onto the sky plane which extends up to about $3 \degr$ from the galactic center and correlates well with the HI velocity field at 21 cm, at least within about $0.5 \degr$ (\citealt{depaolis2016}).
We emphasize that the
very fact that the detected temperature asymmetries are always   almost frequency independent is a 
strong indication of an effect due to the galaxy rotation and remark the importance of
the methodology proposed which, in spite of its simplicity, 
may allow one to consistently estimate the galaxy dynamical
mass contained within a certain galactocentric distance. We have also shown that, in general, our method, can be applied to nearby nearly edge-on spirals and may be used to trace the halo bulk dynamics on rather large scales in a model-independent way.
The present paper is a follow-up of the previous paper on the M82 galaxy (\citealt{gurzadyan2015}) where a  substantial North-South and East-West
temperature asymmetry was found, extending up to about $1 \degr$ from the M82 center. The main conclusion about the origin of the temperature asymmetry (almost frequency-independent) was its link with a 
Doppler-induced effect regarding the line-of-sight dynamics on the real halo scale - invisible in other bands - the ejections from the
galactic center or the tidal interaction of M82 with the M81 galaxy. Here, we consider again this issue from the point of view of the M81 galaxy, one of the last objects in the Local Group which can be studied by available {\it Planck} data. 

\section{Planck data analysis and results toward M81}
M81, also known as Bode's galaxy (or NGC 3031), at J2000 coordinates R.A.: 09$^h$ 55$^m$ 33.1730$^s$, Dec: +69$\degr$ 3$\arcmin$ 55.061$\arcsec$ (Galactic Longitude $l=142.0918406\degr$,  Galactic Latitude $b=40.9001409\degr$) is a SA(s)ab type galaxy at a distance of  $3.6\pm 0.2$ Mpc from us (see, e.g., \citealt{Gerke2011}). Following the procedures described in the previous papers, we have used  the publicly released {\it Planck} 2015 data\footnote{From the {\it Planck} Legacy Archive, http://pla.esac.esa.int.} \cite{planck2015a} in the
bands at 70 GHz of the Low Frequency Instrument (LFI),
and in the bands at 100 GHz, 143 GHz and 217 GHz of the High Frequency
Instrument (HFI).
{We have also used the foreground-corrected SMICA band (indicated as SmicaH in Figs. \ref{fig2} and \ref{fig3}) which should display the lowest contamination by the galactic foregrounds.
We notice here that {\it Planck}'s resolution is 
$13.2\arcmin$, $9.6\arcmin$,  $7.1\arcmin$ and $5\arcmin$ in terms of FHWM at 70, 100, 143 and 217 GHz
bands, respectively, and frequency maps \citep{planck2015b}  are provided in CMB temperature at resolution
corresponding to Nside=2048 in HEALPix scheme \citep{gorski2005}. }
\begin{figure}[h!]
 \centering
  \includegraphics[width=0.46\textwidth]{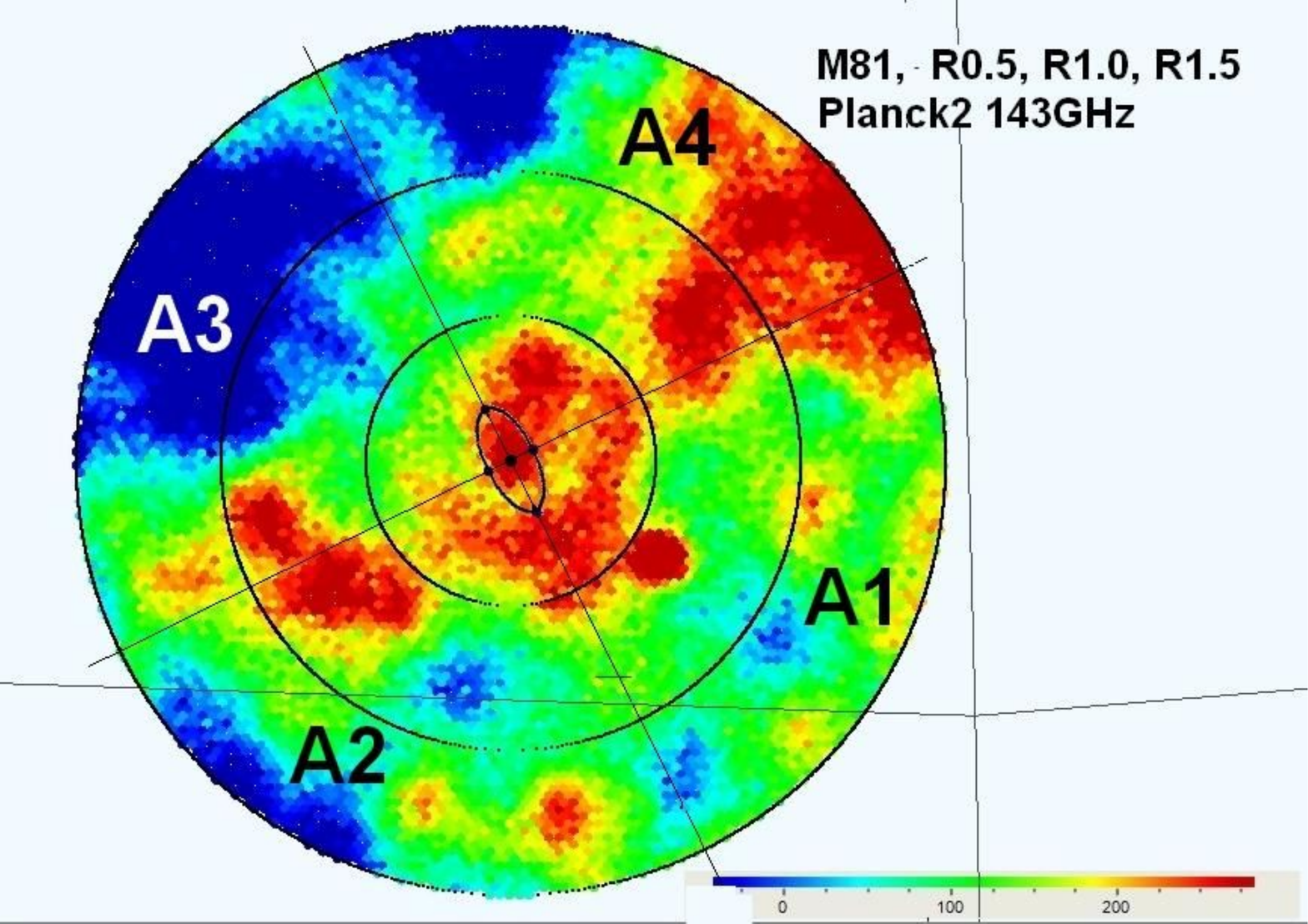}
 \caption{The {\it Planck} field toward the M81  galaxy in the 143 GHz band. The pixel color gives the temperature excess  in $\mu$K with respect to the mean CMB temperature. The optical extension of the M81 galaxy is indicated by the inner ellipse with apparent dimensions   of  $26.9\arcmin$ and $14.1\arcmin$, respectively. The four quadrants A1, A2, A3 and A4 are used in the analysis. The thin dashed black line marks the Galactic latitude $b=40\degr$ North. We note that the M82 galaxy is clearly visible as the red spot in the A1 region at about $38\arcmin$ from the center of M81.} \label{fig1}
 \end{figure}
 \begin{figure}[h!]
 \centering
  \includegraphics[width=0.46\textwidth]{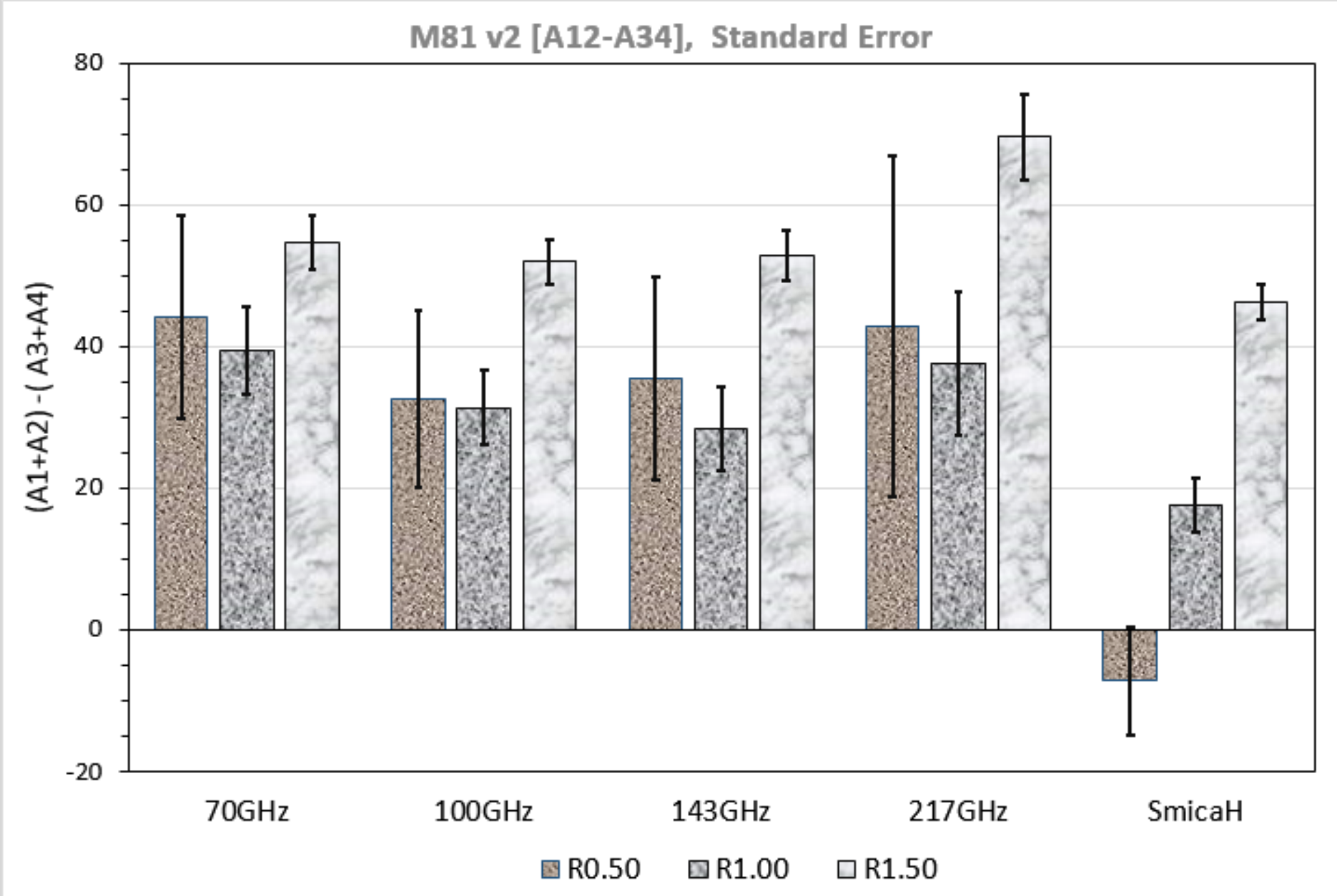}
    \includegraphics[width=0.46\textwidth]{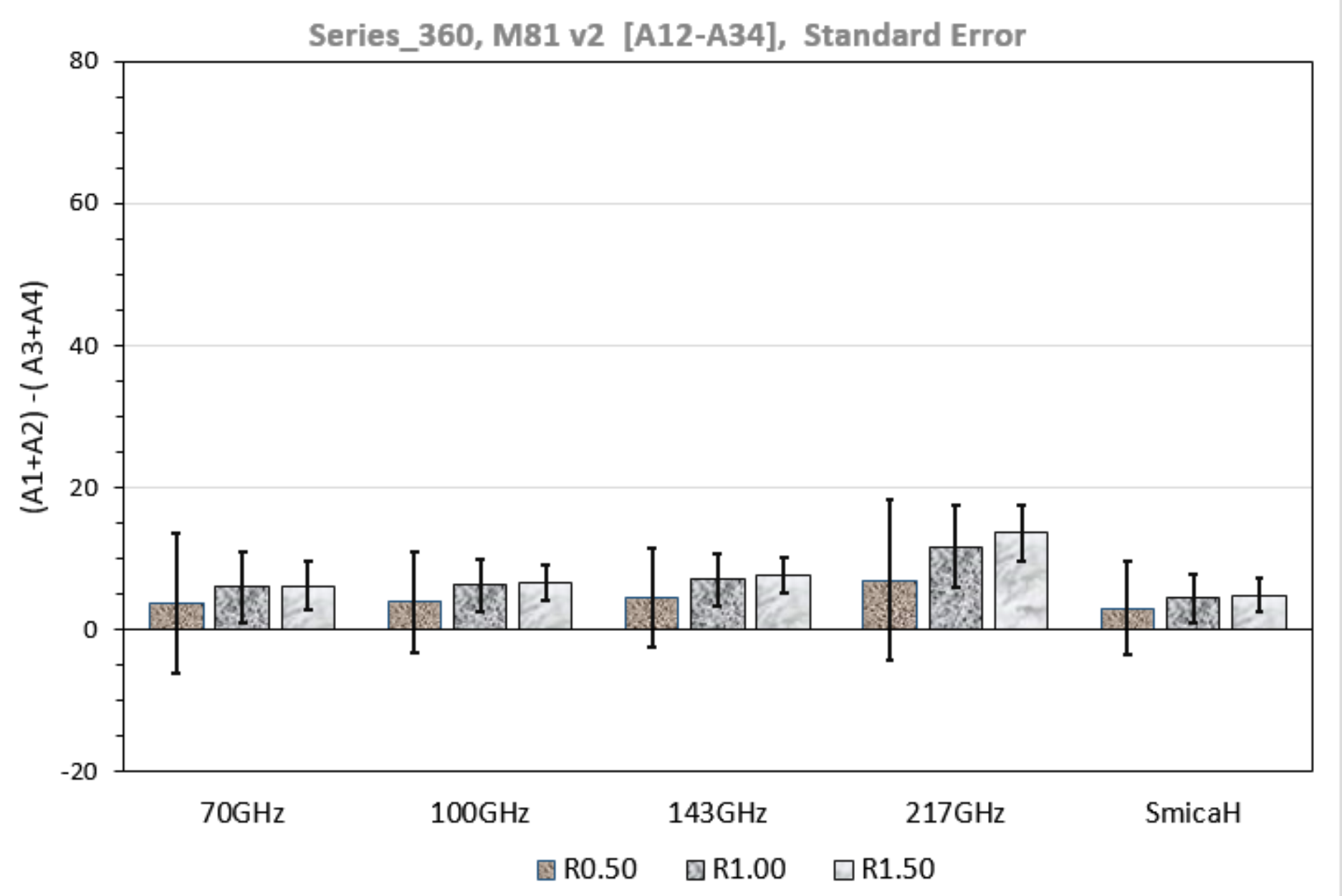}
 \caption{Upper panel: the temperature asymmetry toward M81 in $\mu$K (with the standard errors) of the A1+A2  
region (indicated as A12) with respect to the A3+A4  region (A34)  in the five considered {\it Planck} bands (see text for details) within three radial distances of  $30\arcmin$ ($R0.50$), $60\arcmin$ ($R1.00$) and $90\arcmin$ ($R1.50$). Bottom panel: the same for the 360 control fields with the same geometry (shown in Fig. \ref{fig1}) equally spaced at one degree distance to each other in Galactic longitude and at the same latitude as M81.}
 \label{fig2}
 \end{figure}
   \begin{figure}[h!]
 \centering
  \includegraphics[width=0.46\textwidth]{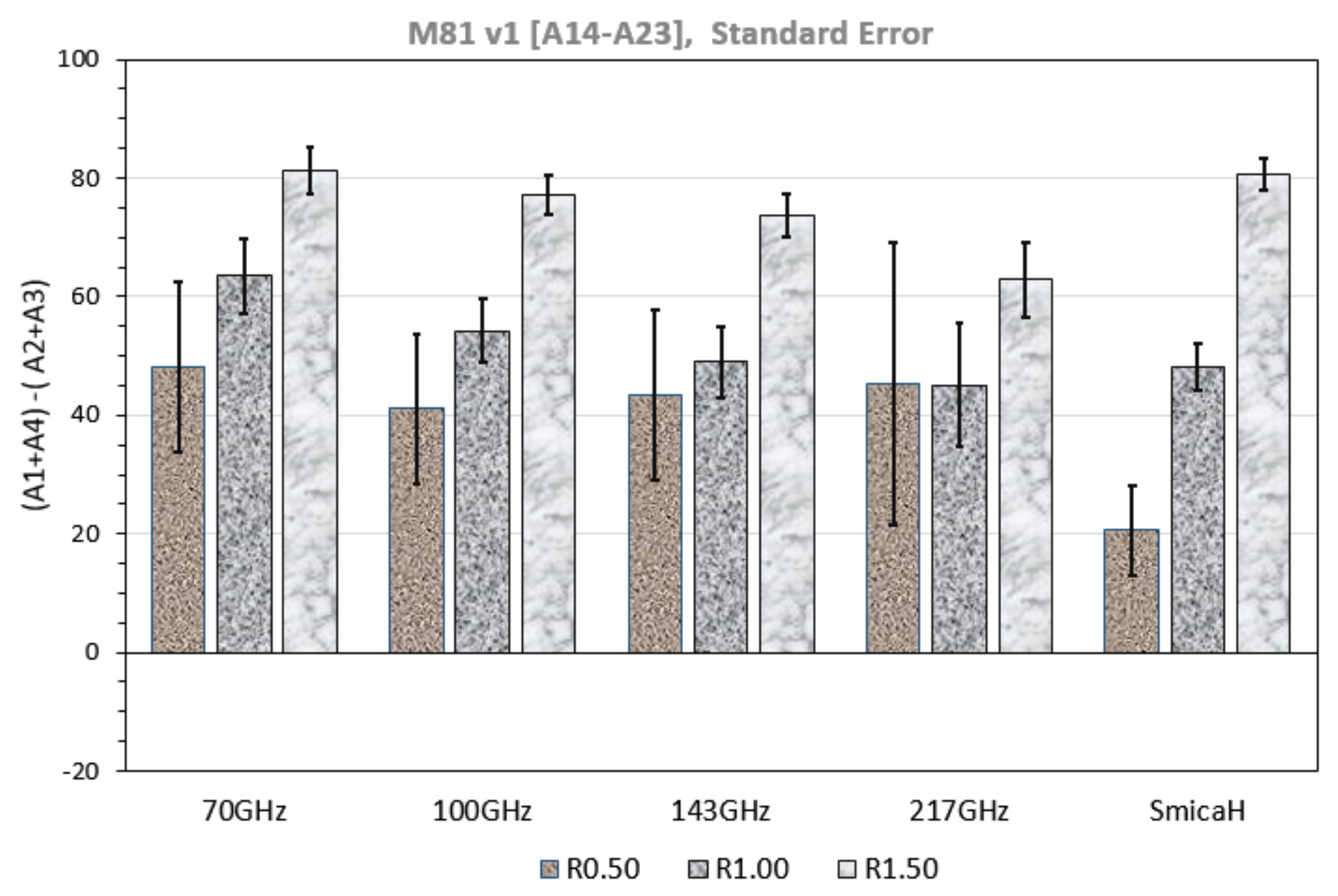}
    \includegraphics[width=0.46\textwidth]{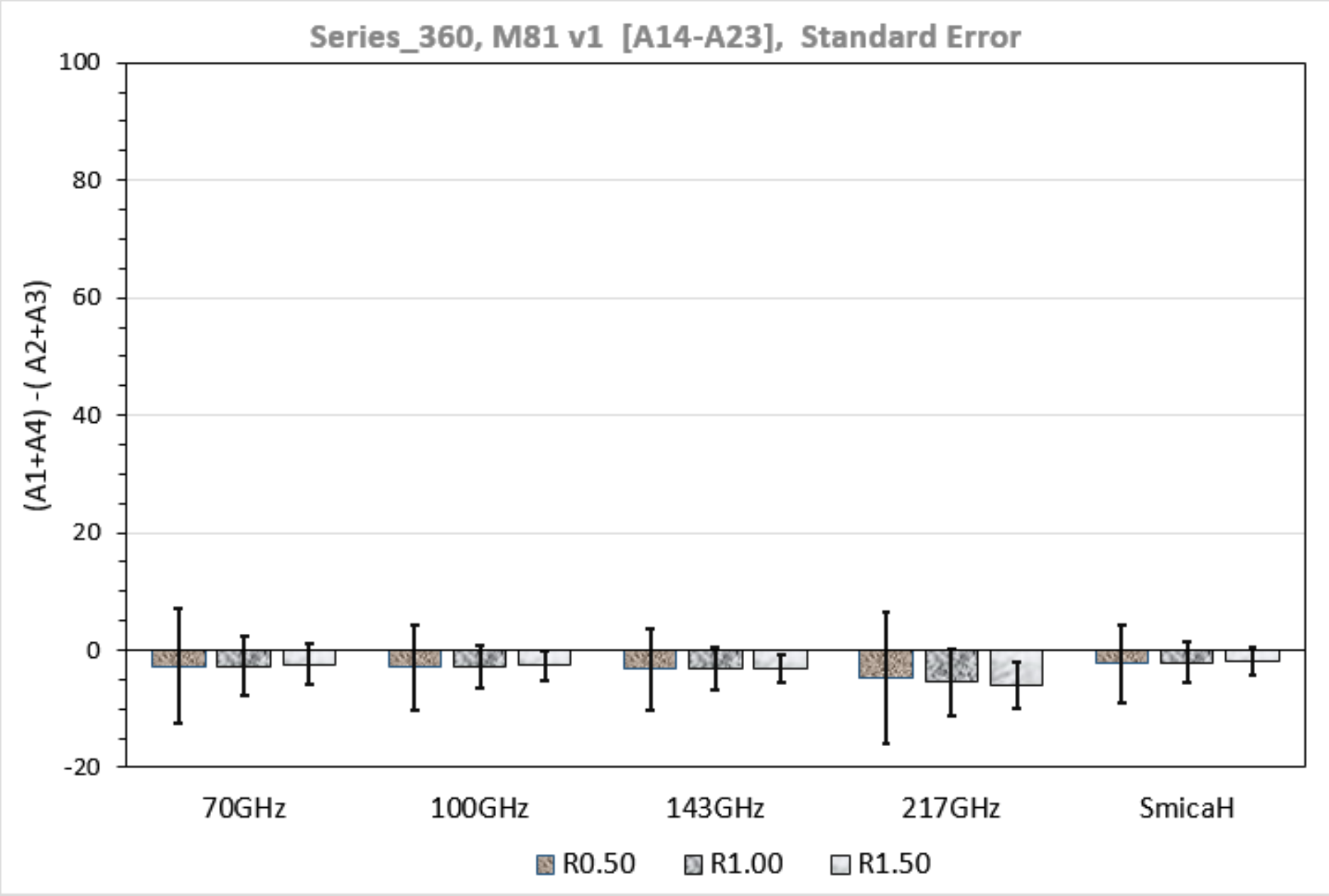}
 \caption{Upper panel: the temperature asymmetry toward M81 in $\mu$K (with the standard errors) of the A1+A4 region (A14) with respect to the A2+A3 region (A23) in the five considered {\it Planck} bands. Bottom panel: the same for the 360 control fields.}
 \label{fig3}
 \end{figure}
{To study in the simplest way the CMB data toward the M81 galaxy, the {\it Planck} field of the region of  interest  (we give in Fig. \ref{fig1} the map in the case of the 143 GHz band) has been divided into  four quadrants A1, A2, A3 and A4. As detailed in the histograms in Figs. \ref{fig2} and \ref{fig3} we have considered the temperature asymmetry in three radial regions about the M81 center within $0.5\degr$, $1\degr$ and  $1.5\degr$ (indicated as  R0.50, R1.00 and R1.50, respectively).
In Fig. \ref{fig1}  the optical extension of the M81 galaxy is  shown, as indicated by the inner ellipse. In Fig. \ref{fig2} we give  the temperature asymmetry toward M81 in 
$\mu$K (with the standard errors) of the A1+A2 region (A12) with respect to the A3+A4 region (A34) in the five considered {\it Planck} bands within the three radial distances. In the bottom panel we give  the same for the 360 control fields with the same geometry (shown in Fig. \ref{fig1}) equally spaced at one degree distance to each other in Galactic longitude and at the same latitude as M81.
As one can see from Fig. \ref{fig2} and as expected by considering the rotation direction of the M81 disk about its rotation axis, The A12 region always appears hotter than the  A34 region by  $32-44~\mu$K within $0.5\degr$, by   $28-40~\mu$K within $1\degr$ and by about  $50-70~\mu$K within $1.5\degr$. }
Note that the control fields show a much smaller temperature asymmetry of only a few $\mu$K and that the detected effect is practically the same in any of the five {\it Planck}'s bands. 
{We also note that the size of the virial radius of the M81 galaxy (equivalent to the $R_{200}$ radius where the galaxy density is about 200 times larger than the critical density) can be estimated to be about $3.6\degr$ \citep{chiboucas2009}.
}
Although from the geometry and the direction rotation of the M81 disk, the expected temperature asymmetry was along the A12/A34 axis, we have also considered  the temperature asymmetry toward M81 in the A14 region with respect to the A23 region and found an even more consistent asymmetry  in all  the considered  {\it Planck} bands. This resembles what was already found towards the companion galaxy M82  \citep{gurzadyan2015}. In the present case the temperature asymmetry amounts to $40-80~\mu$K as implied by a prolate M81 halo rotation, while the control fields always show an asymmetry consistent to zero (see Fig. \ref{fig3}). As far as the 
foreground-corrected SMICA band is concerned, the temperature asymmetry  is negligible within $0.5\degr$ (although one has to consider that the pixel number in this region is very low) and increases to large values within $1\degr$ and $1.5\degr$.  Also SMICA data  show a clear and more consistent  A14/A23 temperature asymmetry  with values comparable with those in the other bands within R1.00 and R1.50, although within $0.5\degr$ there may be a non-negligible foreground contamination in the other {\it Planck}'s bands.


\section{Discussion}
  {Similar to the case of the other galaxies of the Local Group considered previously and in particular toward M82, we found a consistent North-South and East-West temperature asymmetry also toward the M81 galaxy, that reaches values up to  about $80~\mu$K within $1.5\degr$ in all considered {\it Planck} bands.  We believe that the most plausible explanation relies in a Doppler-induced effect due to the spin of the M81 halo, possibly along an axis tilted (up to about $90\degr$) with respect to the rotation axis of the M81 disk. In this case the temperature asymmetry can be estimated from the equation  $ \Delta T/T=2v\sin i S\langle\tau\rangle /c$, accordingly  to the model first discussed in \cite{depaolis1995}, where, $v$ is the M81 rotation velocity, $i\simeq 58\degr$ is the M81 disk inclination angle, $S$ is the gas or dust filling factor, and $\langle\tau\rangle$ is the averaged optical depth within a given {\it Planck} band.  In order to account for the detected temperature anisotropy, however, the M81 halo should be filled by a relatively large  amount of gas (likely in the form of cold gas clouds), as in the models proposed, e.g., by \cite{pfenniger1994,dpijr1995,gs}.
  A viable explanation of the detected effect could  be, in principle, also the rotational kinematic Sunyaev-Zel'dovich (rkSZ) effect, which is known to be active on galaxy cluster scales \citep{cooray2002,chluba2002,manolopoulou2017}. 
 Naturally, to be active, the rkSZ effect does require the presence of a conspicuous population of high energy electrons in the rotating M81 halo.}
  By rescaling to the  free electron density, temperature and size of the galaxy halos, the effect appears to be smaller by at least one order of magnitude (with respect to that in galaxy clusters), giving expected temperature asymmetries below a few $\mu$K. {However, it might be that the temperature asymmetry derives from  a multi component effect due to the presence of cold gas clouds together with a halo of hot and also, possibly, a warm component. Also other effects, such as the case that M81 is an interacting system with a rather recent merging event may induce an increase of the temperature asymmetry, as outlined in the following discussion.

Many galaxies belong to  multiple systems and constitute dynamically linked objects which are affected by their mutual gravitational interaction. An example of such systems  is indeed constituted by the M81, M82  and NGC 3077 galaxies  (the first two objects interacted about 200 Myrs ago,  see e.g. \citealt{makarova2002,sun2005,heithausen2012,oehm2017}). The radio images of the system, obtained for example by the VLA radio telescopes (see \citealt{yun1994}), shows with strong evidence that the HI gas is not only associated with each galaxy but is also present around the galaxies and in the intergalactic space. Clear gaseous filaments are visible  among the three major objects of the galaxy group (which is likely composed by at least six objects: M81, M82, NGC 3077, Holmberg IX, Arp's Loop and NGC 2976 (see, e.g., \citealt{bremnes1998,karachentsev2002,makarova2002}). 
This gas and the associated dust could give some  contribution to the detected temperature asymmetry toward the M81 halo. In this respect we note that an enhancement of the emission arising from the rkSZ effect is expected to occur in recent merging events of rich  galaxy clusters giving temperature asymmetries up to $146~\mu$K \citep{chluba2002}. In the present case, the previous merging event between the M81 and M82 galaxies might have generated large scale turbulence and bulk motion with an increase of the free electron density and temperature which may amplify the rkSZ effect producing possibly temperature asymmetries up to $10-20~\mu$K (although a realistic estimate of the effect would require a detailed hydrodynamic modeling of the past merging event). \footnote{We also mention that presence of baryonic jets from an ultraluminous supersoft X-ray source discovered in the M81 galaxy, called ULS-1 (see \cite{liu2015} and references therein), may lead to an unexpected growth of the high-energy electrons and hot plasma populations in the M81 halo.}

Moreover,  there could  also be some contribution to the detected temperature asymmetry from high-latitude gas clouds in our Galaxy along the line of sight towards M81. In this respect we note that M81 is at about $40.9\degr$ North of the Galactic disk, where contamination from the Milky Way is expected to be low.
However, interpretation of astronomical observations is often hampered by the lack of direct distance information. Indeed, it is often not  easy to judge whether objects on the same line of sight are physically related or not. Since the discovery of the Arp's Loop \citep{arp1965} the nature of the interstellar clouds in this region has been debated; in particular whether they are related  to the tidal arms around the galaxy triplet \citep{sun2005,demello2008} or to Galactic foreground cirrus \citep{sollima2010,davies2010}.
Already \cite{sandage1976} presented evidence showing that we are observing the M81 triplet through wide spread Galactic foreground cirrus clouds and \citealt{devries1987} built large-scale HI, CO and dust maps which showed Galactic cirrus emission towards the M81 region with $N_H\simeq 1-2\times10^{20}$ cm$^{-2}$. The technique used to distinguish between the emission from extragalactic or Galactic gas and dust relies on spectral measurements and on the identification of the  line of sight velocities which are expected to be different in each case. Unfortunately, in the case of the M81 Group, this technique appears hardly applicable since the radial velocities of  extragalactic and Galactic  clouds share a similar LSR (local standard of rest) velocity range \citep{heithausen2012}.
Several small-area molecular clouds (SAMS), i.e. tiny molecular clouds in  a region where the shielding of the interstellar radiation field is too low (so that these clouds cannot survive for a long time) have been detected by \cite{heithausen2002} toward the M81 Group. More recently, data from the SPIRE instrument onboard {\it Herschel} ESA space observatory and MIPS onboard of {\it Spitzer} allowed the identification of several dust clouds north of  the M81 galaxy with a total hydrogen column density in the range $1.5-5\times10^{20}$ cm$^{-2}$ and dust temperatures between 13 and 17 K \citep{heithausen2012}. However, since there  is no obvious difference  among the individual clouds there was no way to distinguish between Galactic or extragalactic origin although it is likely that some of the IR emission both towards M81 and NGC 3077 is of Galactic origin.  Temperature asymmetry studies in {\it Planck} data may be indicative of the bulk dynamics in the observed region provided that other Local (Galactic) contamination in the data is identified and subtracted. This is not always possible, as in the case of the M81 Group, and therefore it would be important to identify and study other examples of dust clouds where their origin, either Galactic or extragalactic, is not clear.\footnote{One such example might be provided by the interacting system toward NGC 4435/4438 \citep{cortese2010} where the SAMS found appear more consistent with Galactic cirrus clouds than with extragalactic molecular complexes.} Incidentally, the region A1 within R0.50 has been studied by \cite{barker2009}, who found evidence for the presence of an extended structural component beyond the M81 optical disk, with a much flatter surface brightness profile, which  might  contain $\simeq 10-15\%$ of the M81 total V-band luminosity. However the lack of both a similar analysis in the other quadrants (and at larger distances from the M81 center) and the study of the gas/dust component associated to this evolved stellar population, hamper understanding whether  this component may explain the observed temperature asymmetry toward the M81 halo.

Although  the physical origin  of the detected  temperature asymmetry is not clearly identified at present, it appears obvious that the CMB asymmetry method is tracing the M81 halo and intergalactic bridges, not directly revealed in other bands and via conventional methods, based on stellar population studies, ISM, etc. In order to assess this issue, high-resolution and extended (up to $\simeq 1.5 \degr$) observations to infer the distribution of the cold, warm and hot gas components  appears to be an urgent task to be performed. In this respect we emphasize that, in addition to radio observations at 21 cm to map the HI component of the gas (integrated also by other techniques to study small-scale cold structures as done, e.g.,  through interstellar scintillations by \citealt{habibi2011}) and to the X-ray band diffuse emission to infer the amount and distribution of the hot gas component, investigation of the warm gas component with the methodology employed, e.g.,  in \cite{nicastro2016} is extremely important.
Given the serious quantitative disagreement between the microwave temperature asymmetry amplitude revealed for M81 and several other  nearby galaxies and the rkSZ contributions there, the latter's alternative may be  more exotic  halo models (see, e.g., \citealt{Lo2017,Ok2017,Pi2017,van2017,GK}), a dilemma to be solved by future studies. 

In conclusion, the available {\it Planck} data, by now, enabled one to trace, by this method, the haloes in the nearby edge-on spirals previously analyzed, while higher resolution data can be efficient for the studies of galaxies also outside the Local Group.  
This is particularly important in view of the next generation CMB experiments, such as LiteBird \citep{LiteBird}, CMB-S4 \citep{CMB-S4}, CORE \citep{CORE}, DeepSpace\footnote{See the DeepSpace website at http://deep-space.nbi.ku.dk.}, PIXIE \citep{PIXIE},  and Polarbear \citep{Polarbear}, which will attempt even more precise measurements of the CMB than available so far. Many of these experiments are designed to cover mainly the frequency range around 100 GHz where the relative intensity of the CMB is known to be highest and where one of the most dominant foreground components is dust emission (see, e.g.,
 \citealt{liu2017}). Understanding the properties of dust emission and distinguishing between Galactic foregrounds and extragalactic emission is an important premise for the optimized use of the next generation CMB experiments.

 

\begin{acknowledgements}
We acknowledge the helpful comments by the referee and the use of {\it Planck} data in the Legacy Archive for
Microwave Background Data Analysis (LAMBDA) and HEALPix
\citep{gorski2005} package. FDP, AAN, and GI acknowledge the support by the INFN
projects TAsP and EUCLID. PJ acknowledges support from the
Swiss National Science Foundation. 
AQ is grateful for hospitality to the DST Centre of Excellence in Mathematical \& Statistical Sciences of the University of the Witwatersrand, Johannesburg. 
\end{acknowledgements}


\end{document}